\documentclass[twocolumn,preprintnumbers,notitlepage,superscriptaddress,amsmath,10.5pt,aps,prl]{revtex4-1}
\usepackage{graphicx}
\usepackage{color}
\usepackage{hyperref}
\usepackage{subfigure}
\usepackage{graphicx}
\usepackage{type1cm}
\usepackage{eso-pic}
\usepackage{color}
\usepackage{amsmath}
\usepackage{amssymb}
\usepackage{textcomp}
\usepackage[T1]{fontenc}

\def\bluee#1{\textcolor{black}{#1}}

\begin{document}

\title{An All-dielectric Metasurface Polarimeter}


\author{Yash D. Shah}
\thanks{These authors contributed equally to this work.}
\affiliation
{School of Physics and Astronomy, University of Glasgow, Glasgow, UK}

\author{Adetunmise C. Dada}
\thanks{These authors contributed equally to this work.}
\affiliation
{School of Physics and Astronomy, University of Glasgow, Glasgow, UK}

\author{James P. Grant}
\affiliation
{Microsystems Technology Group, James Watt School of Engineering, University of Glasgow, Glasgow, UK.}

\author{David R. S. Cumming}
\affiliation
{Microsystems Technology Group, James Watt School of Engineering, University of Glasgow, Glasgow, UK.}

\author{Charles Altuzarra}
\affiliation
{School of Physics and Astronomy, University of Glasgow, Glasgow, UK}

\author{Thomas S. Nowack}
\affiliation
{Microsystems Technology Group, James Watt School of Engineering, University of Glasgow, Glasgow, UK.}

\author{Ashley Lyons}
\affiliation
{School of Physics and Astronomy, University of Glasgow, Glasgow, UK}

\author{Matteo Clerici}
\affiliation
{James Watt School of Engineering, University of Glasgow, Glasgow, UK.}

\author{Daniele Faccio}
\affiliation
{School of Physics and Astronomy, University of Glasgow, Glasgow, UK}

\email{Daniele.Faccio@glasgow.ac.uk}

\begin{abstract}
 The polarization state of light is a key parameter in many imaging systems. For example, it can image mechanical stress and other physical properties that are not seen with conventional imaging, and can also play a central role in quantum sensing. 
  However, polarization is more difficult to image and polarimetry typically involves several independent measurements with moving parts in the measurement device. Metasurfaces with interleaved designs have demonstrated sensitivity to either linear or circular/elliptical polarization states. 
  Here we present an all-dielectric meta-polarimeter for direct measurement of any arbitrary polarization states from a single unit-cell design. By engineering a completely asymmetric design, we obtained a metasurface that can excite eigenmodes of the nanoresonators, thus displaying a unique diffraction pattern for not only any linear polarization state but all elliptical polarization states (and handedness) as well. 
 The unique diffraction patterns are quantified into Stokes parameters with a {resolution of 5$^{\circ}$ and with a} polarization state fidelity of up to $99\pm{1\%}$. 
 This holds promise for applications in polarization imaging and quantum state tomography.
\end{abstract}

\maketitle

\section{Introduction}
Metasurfaces or flat-optics~\cite{Chen2020,Yu2014} are the two-dimensional counterparts of metamaterials comprising of subwavelength-thickness nanostructures. The spatial distribution, geometry and material (metallic or high-index dielectric) of these nanostructures provide dramatically enhanced light-matter interactions at subwavelength scales. This can be utilized to efficiently control and tailor local polarizations, phases and amplitudes  of linear fields~\cite{BalthasarMueller2016,Basiri2019}. Metasurfaces have emerged as a promising technology for unprecedented manipulation of light,  which makes them integral for nanophotonics and miniaturization of optical systems. 
Polarimetry, i.e. the ability to \bluee{measure one} of the fundamental properties of light---polarization---is of significant importance from research in fundamental physics and light-matter interaction to polarization-based quantum information technology and quantum imaging~\cite{Altuzarra2019,Wang2018,Solntsev2021}.  From polarimetry we typically obtain the Stokes parameters which contain all the information to represent the State of Polarization (SoP) of light~\cite{Georgi2019}. Whilst traditional polarimetry requires multiple measurements, even with moving or reconfigurable systems~\cite{tyo2006review}, metasurfaces hold promise for measurements based on static and ultra-compact components. The constituent elements that make up a metasurface (meta-atoms) can be engineered to form subwavelength-scale polarization optics with metal-based (plasmonics)~\cite{Stokes1852} or dielectric-based material systems~\cite{Ding2018}. 
Polarization-sensitive metasurfaces are realized predominantly with anisotropic nanopillars (elliptical or rectangular). The orientation of these nanostructures lead to birefringence and sensitivity to the linear polarization state of light. The Pancharatnam-Berry (PB) phase or geometric phase is the accumulated phase due to the orientation and arrangement of \bluee{anisotropic} meta-atoms, making it sensitive to circularly polarized light. Thus, various designs have been realized for plasmonic polarimeters~\cite{Intaravanne2020,Ding2018}. However, a major drawback is the high absorption loss inherent in metallic structures. On the other hand, the use of dielectric-based material systems has shown improved performances with higher diffraction (and transmission) efficiencies. Forming nanostructure resonators from materials with a high refractive index contrast leads to higher scattering of electromagnetic fields, and combining \bluee{this} with engineering of the geometry and spatial distribution leads to higher efficiencies for SoP measurement. 

For direct measurements of a polarization state, the geometry (which relates to propagation phase) and orientation (which relates to geometric phase) of anisotropic nanopillars \bluee{are judiciously} varied providing unique designs for each \bluee{of the horizontal ($\left | H  \right \rangle$), vertical ($\left | V  \right \rangle$), anti-diagonal ($\left | A  \right \rangle$), diagonal ($\left | D  \right \rangle$), left-circular ($\left | R  \right \rangle$) and right-circular ($\left | L  \right \rangle$)} polarization states, which are then interleaved into one metasurface platform~\cite{Bai2019}.  While this approach could differentiate between these 6 \emph{degenerate polarization states}, \bluee{it has difficulty measuring} other polarization states, leading to spurious diffraction at the point of interlacing. By imposing arbitrary propagation and geometric phase profiles with birefringent rectangular nanopillars, designs could address elliptical polarization states as well~\cite{Arbabi2018,BalthasarMueller2017}.  Extending this approach and using the Fourier transform of the optical field for the Jones matrix of each element, elliptical and circular polarization states were diffracted to specific spots and used for polarization-based imaging~\cite{Rubin2018}. However, in this case a design that is sensitive to linear polarization states would have the same response to orthogonal elliptical/circular states.  Holography provides a visual approach to polarimetry wherein a unique hologram needs to be encoded using the phase information of each meta-atom for each polarization state~\cite{Rubin2019}. This makes it difficult to differentiate between arbitrary polarization states. The holographic design proposed in Refs.~\citenum{Song2020,Zhang2019}  extracts the Stokes parameters  however, with a limitation in resolution as the periodicity limits how many designs can be incorporated in a single unit cell. \bluee{A} single-cell design that can be designed at any desired wavelength and can successfully distinguish between not only linear and elliptical but also opposite handedness has not been realized.

\begin{figure*}[htbp]
  \centering
  \includegraphics[width=0.8\textwidth]{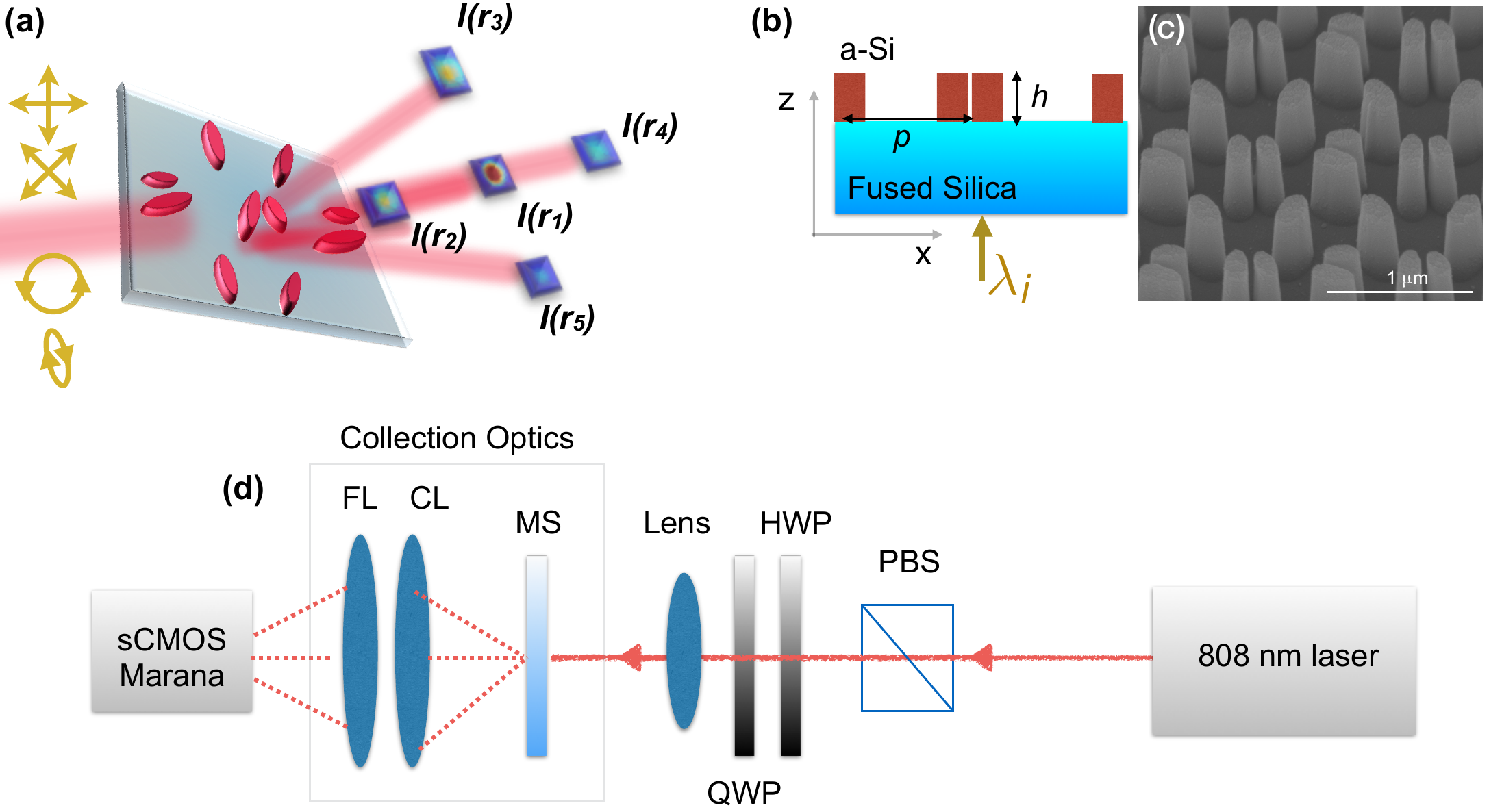}
\caption{(a) Schematic of the proposed meta-polarimeter which gives a diffraction pattern unique to any arbitrary incident polarization state. $I(r_n)$ is the intensity at each diffracted spot $n$. (b) Schematic of the cross-section with light ($\lambda_i$=808 nm) incident from the substrate. (c) SEM image of the fabricated metasurface. (d) Schematic of the measurement setup to obtain a single-shot diffraction pattern on a Andor Marana sCMOS camera. PBS: Polarizing beam splitter. HWP: Half-wave plate. QWP: Quarter-wave plate. MS: metasurface. CL: Condenser lens. FL: Fresnel lens. }
\label{Figure1}
\end{figure*}

Here we introduce an all-dielectric metasurface design comprising of elliptical nanopillars that, with a single unit cell design, provides a unique diffraction pattern for any input polarization state, covering the entire Poincaré sphere. Unlike the metasurface designs based on the PB phase, our approach relies on the control of the resonant modes supported by the nanopillar, which affects the electric and magnetic multipoles.  While plasmonic metasurfaces exploit  electrical resonances, dielectric nanostructures are the ideal choice as they enhance not only electrical, but also magnetic resonances. Our asymmetric design exploits exotic multipolar coupling by exciting otherwise symmetry-protected eigenstates, which makes the proposed design sensitive to not only linear, but circular/elliptical polarized light as well.

\begin{figure*}[htbp!]
  \centering
  \includegraphics[width=0.8\textwidth]{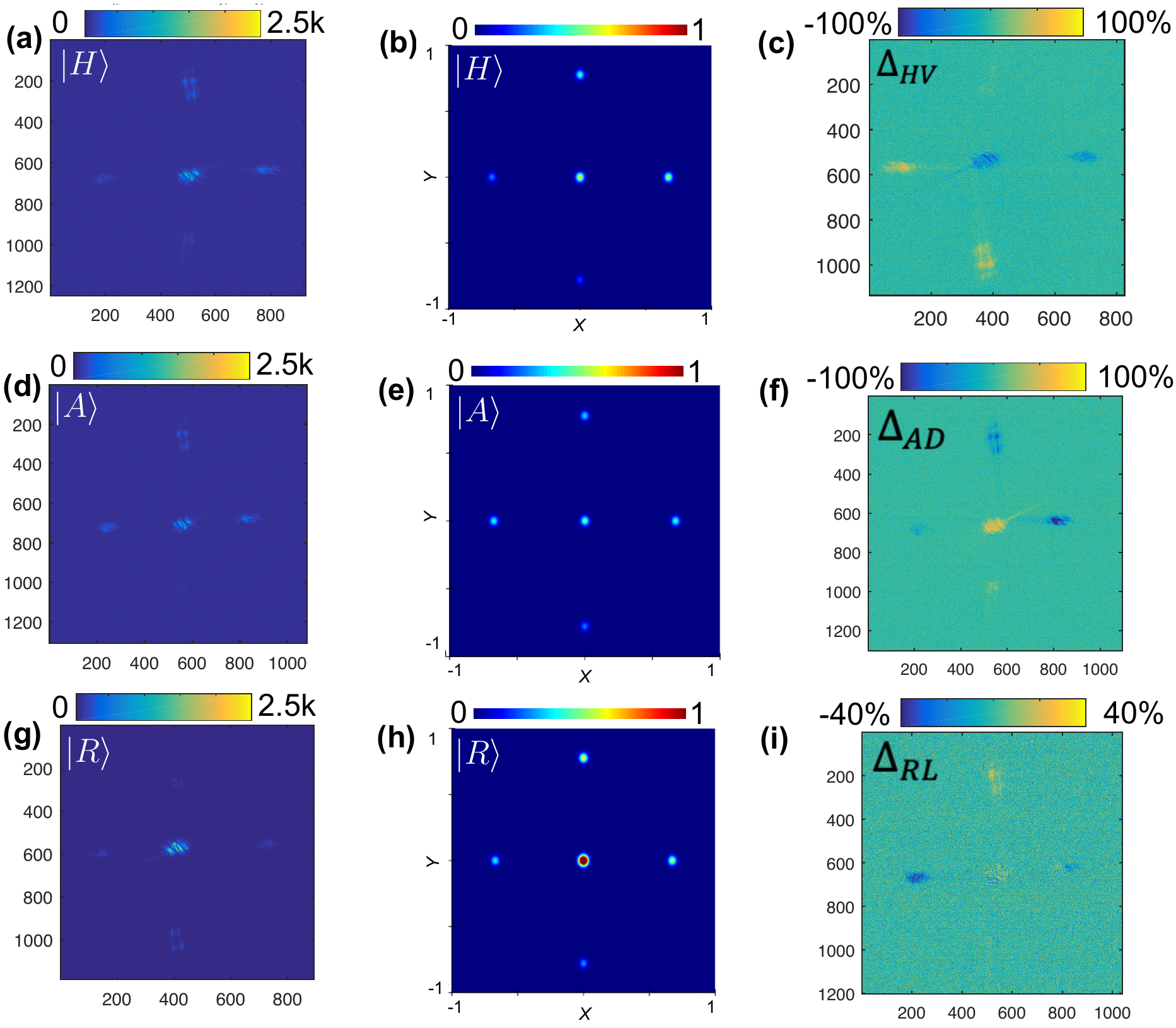}
\caption{Experimental diffraction patterns and \bluee{finite-difference time-domain} (FDTD) simulations for the orthogonal polarization states. The experimental diffraction spots are plotted with the scale indicating the counts and the simulated results are plotted with the scale normalized to $|E|^2$ in the far field for (a) and (b) $|H\rangle$ polarization state, (d) and (e) $|A\rangle$ polarization state and (g) and (h) $|R\rangle$  polarization state, respectively. To highlight the \bluee{differences between intensity patterns for} orthogonal states, (c), (f) and (i) show the $\%$ difference in the intensities of the diffracted spots between $|H\rangle$ and $|V\rangle$ ($\Delta_{HV}$), $|A\rangle$ and $|D\rangle$  ($\Delta_{AD}$) and $|R\rangle$ and $|L\rangle$ ($\Delta_{RL}$), respectively. \bluee{Horizontal and vertical axes represent the respective positions within the lateral-plane image.} 
}
\label{Figure2}
\end{figure*}

\bluee{Our design is based on pairs of elliptical pillars which are combined to form the unit cells of the metasurface, which are referred to here as \emph{bi-meta-atoms}. These are arranged} such that, for any input polarization of light with wavelength $\lambda_i\approx$  810 nm normally incident  on the metasurface, we obtain 5 diffraction spots of varying intensity $I(r_n)$ with a unique distribution depending on the polarization state, as shown in Fig.~\ref{Figure1} (a). Light is incident normally on the fused-silica substrate  and the amorphous-silicon ($\alpha$-Si) elliptical nanopillars, see Fig.~\ref{Figure1} (b). The SEM image of the metasurface is shown in Fig.~\ref{Figure1} (c). 
A period $p=600$ nm with { $p>2\lambda/\mu$ (where $\mu$ is the refractive index \bluee{of the elliptical pillars})} excites diffraction modes of orders $\pm1,0$ 
at an angle $\theta \approx \sin^{-1}(\lambda/p$).  
The design principle was based around finding a range of values for geometric parameters, {including the} orientation of the elliptical pillars and height ($h$), for which we obtain not only the largest difference \bluee{in the detected intensity distribution} between $|L\rangle$  and $|R\rangle$  states, but also high transmission (thereby higher diffraction efficiency). From finite-difference time-domain (FDTD) simulations, the height $h$ of the nanopillars was selected as $520$ nm which gave the highest transmission efficiency \bluee{of up to 63.7\%} (see the Section S1 of the Supplementary Information). Extensive simulations were performed for the geometry and orientation of the elliptical pillars (see Sections S1 and S3 of the Supplementary Information). A ratio of 0.5 is selected between the \bluee{minor and major} axes of the ellipse \bluee{to accommodate} resonances at different wavelengths without overlap. Post fabrication, the devices were characterized using the setup shown in Fig.~\ref{Figure1} (d) to obtain a single-shot image of the diffraction pattern using a condenser lens (CL) and \bluee{a 2-inch-diameter} Fresnel lens (FL) (see Methods for details). The diffraction patterns for the six \bluee{degenerate} polarization states, $|H\rangle$, $|V\rangle$, $|A\rangle$, $|D\rangle$, $|R\rangle$, and $|L\rangle$ are measured on a CMOS camera (Andor Marana) and examples are  shown in Fig.~\ref{Figure2} along with the corresponding FDTD  simulation results (see Methods).
\section{Results and discussion}
\subsection{Experimental results}
Each polarization state has a unique diffraction pattern with different intensity, $I(r_n)$, in the spatially distributed spots, $n=1,\ldots,5$. Figures~\ref{Figure2} (a), (d) and (g) show the diffraction patterns from $|H\rangle$, $|A\rangle$ and $|R\rangle$ polarization states respectively, along with the simulated results in Figures~\ref{Figure2} (b), (e) and (h). We also plot the intensity difference ($\Delta I$) of each spot for the orthogonal basis polarization states in Figs.~\ref{Figure2} (c), (f), (i). The stark differences showcase how the diffraction patterns allow us to distinguish not only  $|H\rangle$ ($|A\rangle$) from $|V\rangle$ ($|D\rangle$), but also the handedness of circular polarization, $|R\rangle$ and $|L\rangle$.
%

\begin{figure*}[htbp!]
\centering
\includegraphics[width=1\textwidth]{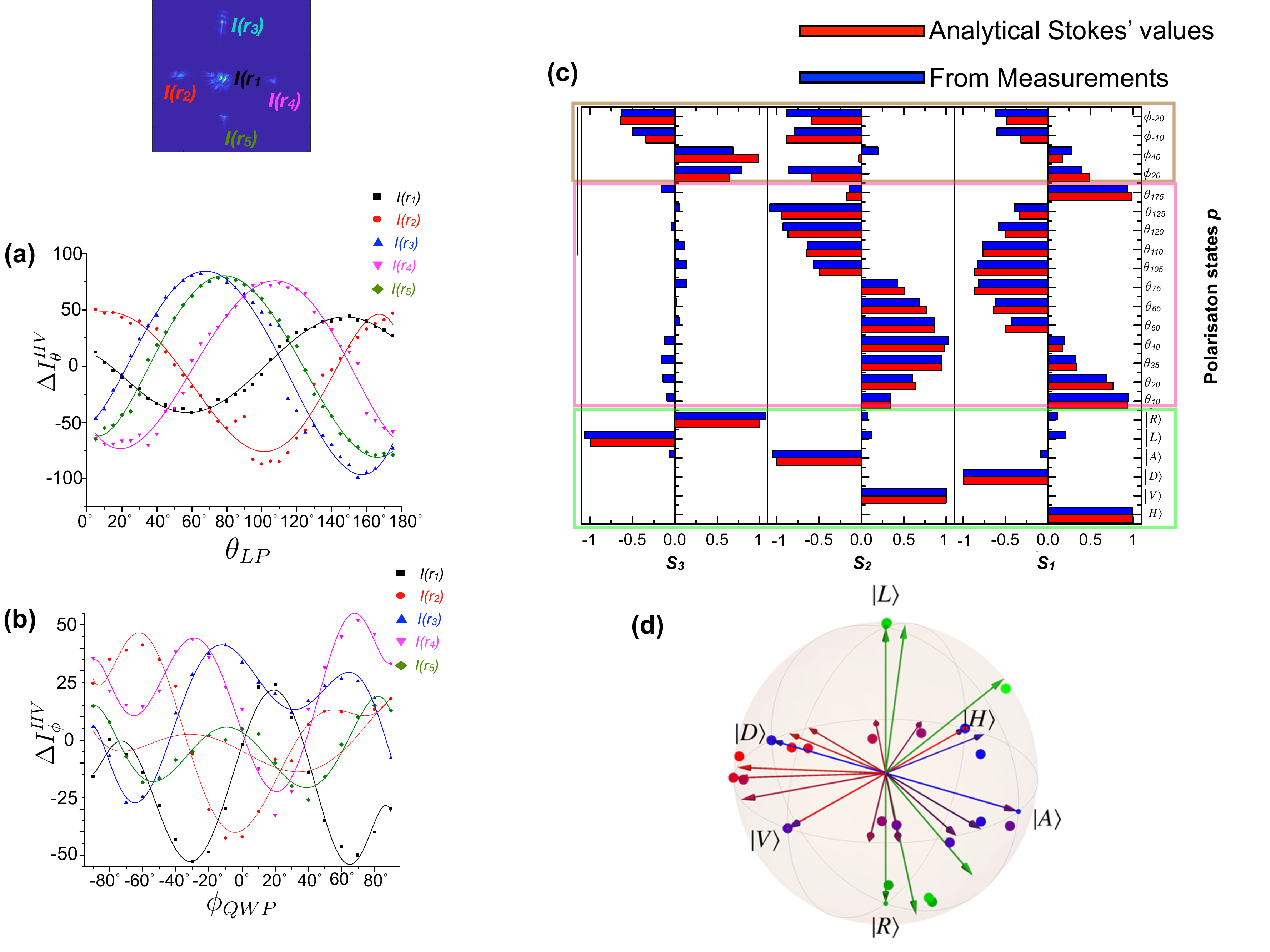}
\caption{Using the intensity of the diffracted spots to extract the Stokes parameters and comparing with analytical Stokes’ values (a) The difference in intensities, $\Delta I^{HV}_{\theta}$, of the 5 diffraction spots for all linear polarization states, $\theta_{LP}$. (b) The difference in the intensities of the 5 spots, $\Delta I^{HV}_{\phi}$ for all elliptical polarization states corresponding to QWP angle $\phi_{QWP}$ with HWP maintained at $22.5^{\circ}$. (c) Comparison of the experimental Stokes parameters for various polarization states obtained using the metasurface matrix (blue) and analytical Stokes’ values (red). The \bluee{degenerate polarization} states are highlighted in green, the linear polarizations in pink and the elliptical polarizations in brown. (d)  Representation as points on the Poincaré sphere of some of the input states (arrows) and the corresponding states experimentally identified by the metasurface (large dots) from results shown in (a)-(c). These measurements have an average fidelity of 99.27$\pm$0.86$\%$.}
\label{Figure3}
\end{figure*}

In Fig.~\ref{Figure3} (a), we show $\Delta I_{\theta}^{HV}(r_n)=\left ( I_{rn}^{HV}-I_{rn}^{\theta}\right )/I_{rn}^{HV} $, where $I_{rn}^{\theta}$ is the intensity for the linear polarization input state defined by angle $\theta$, and $I_{rn}^{HV}$ is the sum of the respective intensities for input states $|H\rangle$ and $|V\rangle$, for the $n$th diffraction spot. 
Similarly, Fig.~\ref{Figure3} (b)  plots $\Delta I_{\phi}^{HV}(r_n)=\left ( I_{rn}^{HV}-I_{rn}^{\phi}\right )/I_{rn}^{HV}$, where $I_{rn}^{\phi}$ is the intensity for the elliptical polarization input state defined by angle $\phi$ (with the HWP fixed at $22.5^{\circ}$ and the QWP at angle $\phi$), for the $n$th diffraction spot.

We then need to {determine} the Stokes parameters~\cite{Rubin2021} from these intensity patterns. From the intensity difference plots in Fig.~\ref{Figure3} (a), we observe that the intensity trends of the diffracted spot at position $r_3$ and $r_5$ {are} nearly equal. In other words, we need only consider $r_3$ or $r_5$ but not both. Thus, the information from the intensity trends of diffracted spots at the four positions $r_1$, $r_2$, $r_4$ and $r_5$ is used to obtain the Stokes parameters, $S_1$, $S_2$ and $S_3$. 

The intensity from various spots from all polarization states can be quantified by a metasurface matrix, $[M]$, such that $[I] = [M]\times[S]$; where $[I]$ is the intensity  vector and $[S]$ is the Stokes  vector. Explicitly,
\begin{equation}
\begin{bmatrix}
I_1\\ 
I_2\\ 
I_4\\ 
I_5
\end{bmatrix}=\begin{bmatrix}
M_{11} & M_{12} & M_{13}\\ 
M_{21}& M_{22} &M_{23} \\ 
 M_{31}& M_{32} &M_{33} \\ 
 M_{41}& M_{42} & M_{43}
\end{bmatrix}\begin{bmatrix}
S_1\\ 
S_2\\ 
S_3
\end{bmatrix}.
\end{equation}
The intensity vector $[I]$ contains the intensity differences either  $\Delta I_{\theta} ^{HV}$ or $\Delta I_{\phi}^{HV}$ and the Stokes vector $[S]$ contains the Stokes parameters ($S_1$, $S_2$, $S_3$) normalized to $S_0$. The metasurface matrix $[M]$ is calculated from the intensities of the $|H\rangle$, $|V\rangle$, $|A\rangle$, $|D\rangle$, $|R\rangle$, and $|L\rangle$ states. We note that $[M]$ embodies the linear relationship between the Stokes parameters and the elements of the intensity vector~\cite{Wei2017}. We determined $[M]$ to be 
\begin{equation}
  M=\begin{bmatrix}
0.2251 & -0.3462 & 0.2374\\ 
 0.3688& 0.0753 &-0.2374 \\ 
 -0.3841& 0.4570 &-0.2464 \\ 
 -0.2662& -0.0714 & 0.1039
\end{bmatrix}
\end{equation}
which can be shown to be a matrix of rank 3, indicating the ability to span the full 3 dimensional vector space defined by the Stokes parameters ($S_1$, $S_2$, $S_3$).  \bluee{As such, it is able to map any polarization state} on the Poincaré sphere, a necessary condition in order to perform polarimetry.
 Using this matrix and the measured intensities observed from arbitrary polarization states, any input polarization can be identified from the corresponding Stokes parameters.  Figure~\ref{Figure3} (c) shows a good match between the Stokes parameters obtained from the measured intensities \bluee{and those corresponding to the input polarization states}~\cite{Rubin2021,Collett2005}. From the graphs and the extracted Stokes values, with this meta-polarimeter we are able to resolve not only any linear polarization states, but also elliptical polarizations and can \bluee{determine any} arbitrary polarization with a resolution of $5^{\circ}$  (see Section S2 of the Supplementary Information for more details). We calculated the fidelities between the polarization states measured using the meta-polarimeter  $\left |  \Psi _{\text{meas}} \right \rangle$ and the respective input polarization states $\left |  \Psi _{\text{in}} \right \rangle$ as $F=  \left |\left \langle   \Psi _{\text{meas}}|\Psi _{\text{in}}\right \rangle  \right |^2$. The input and measured states plotted on the Poincare sphere in Fig.~\ref{Figure3} (d) give an average fidelity of F = $99.27\pm0.86\%$. A clear 
 difference in the intensities of various spots is seen in the diffraction patterns for the $|L\rangle$/$|R\rangle$ and elliptical polarizations as shown in   Figure~\ref{Figure3} (c) and from analysis in Figure~\ref{Figure3} (d).

In terms of understanding the underlying physics, the polarization effect of a metasurface has been described by a Jones matrix that can be written in terms of  the eigenmodes supported in the nanoresonator, $\Psi ^\pm_{eig}$~\cite{Shi2020}; 
\begin{equation}
J=\begin{bmatrix}
E_x\\ 
E_y
\end{bmatrix}=e^{i\phi}\left ( e^{\frac{i\Delta}{2}}|\Psi ^+_{eig}\rangle\langle\Psi ^+_{eig}| +e^{\frac{-i\Delta}{2}}|\Psi ^-_{eig}\rangle\langle\Psi ^-_{eig}|\right )
\end{equation}
where $\phi$ and $\Delta$ are the phase and phase retardation terms. Engineering the meta-atoms to excite different eigen-polarization states is the principle focus of this design. Given the asymmetric nature of the proposed design, we are not limited to bound states 
as is the case for the designs explored so far in literature. Another degree of freedom exploited is that, for the bi-meta-atoms, the resonant modes~\cite{Leitis2019,Koshelev2018} occur in the pillars as well as the gap between the pillars. The different resonant eigenmodes that are excited by the different polarization states of the incident light are a combination of electric dipole (ED), magnetic dipole (MD), electric quadrupole (EQ) and magnetic octupole (MO) {modes}. The modes are a combination of Mie  mode (surface) and the Fabry Perot mode formed within the pillar (open-ended oscillator approximation)~\cite{Kuznetsov2016,Kivshar2018,Koshelev2020}. The interplay between the vector moments of the various poles excited by the polarization of light results in a unique electric and magnetic ($E$-$H$) field patterns. Investigating the eigenmodes excited by the linearly polarized incident light, we analyzed the simulated normalized $E$-$H$ fields of $|A\rangle$ polarization state  from the bi-meta-atom arrangement highlighted in the schematic  in Fig. \ref{Figure4}. (Section S3 of the Supplementary Information contains the $E$-$H$ fields for $|H\rangle$, $|V\rangle$,   $|A\rangle$, and $|D\rangle$ polarization states).  

\begin{figure*}[htbp!]
  \centering
  \includegraphics[width=0.8\textwidth]{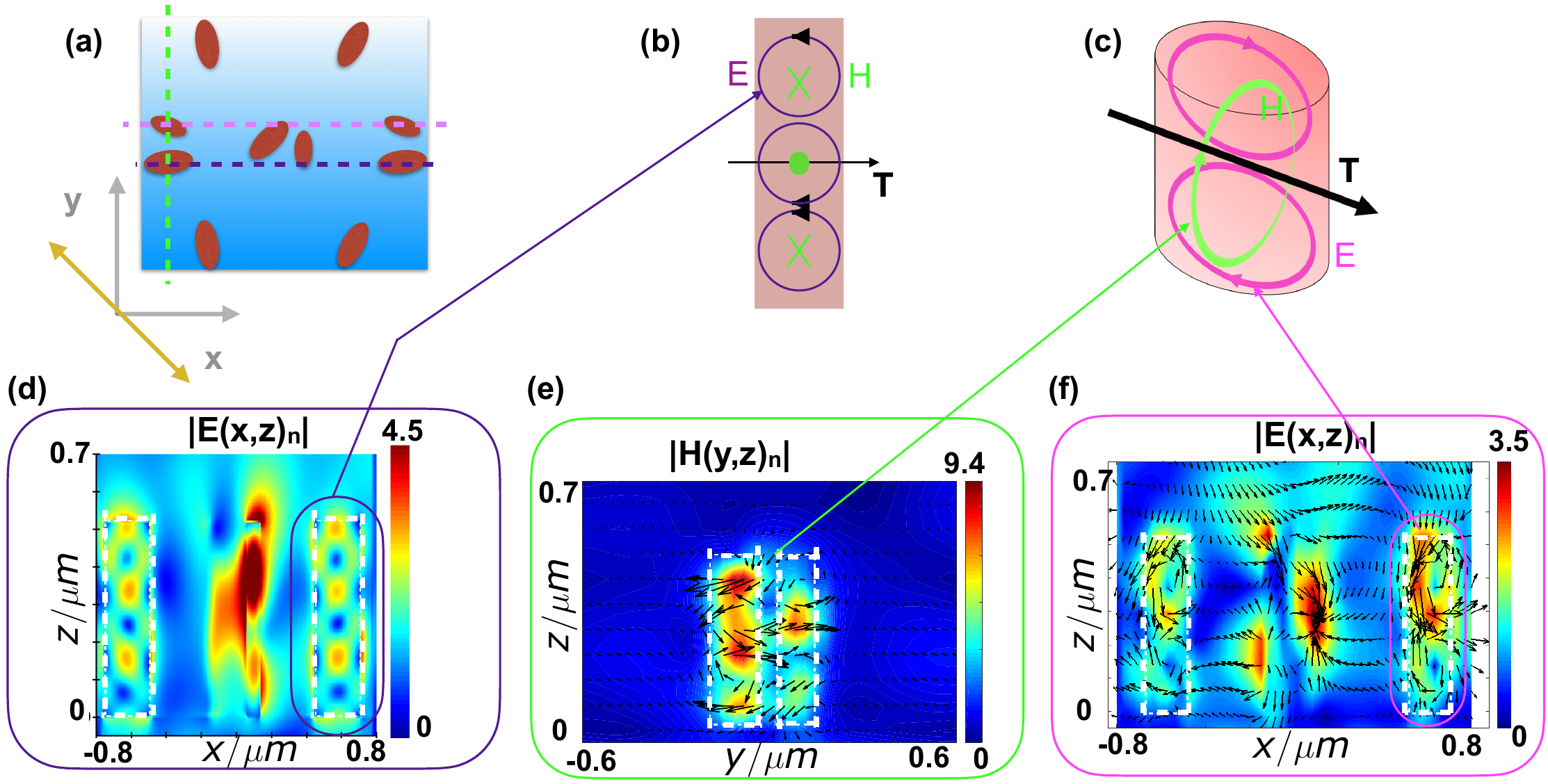}
\caption{
Normalized $|E|$ and $|H|$ fields of the bi-atom arrangements for $|A\rangle$ polarization of incident light. (a) Illustration of the unit cell and its orientation. (b) Illustration of the resonant modes.  (c) Illustration of the anapole showing the coupling between the $E$ and $H$ fields. (d) The normalized E-field $|E(x,z)_n|$ for the larger meta-atom shown in the schematic. (e) The normalized $|H(y,z)_n|$ field and the vector field for the smaller meta-atom. (f) The signature of the formation of a non-radiating anapole state as observed from the simulated E-field (highlighted).  
}
\label{Figure4}
\end{figure*}

Figure \ref{Figure4} shows the simulated $|E(x,z)_n|$ and $|H(y,z)_n|$ of the bi-meta-atoms arrangement as indicated in the schematic for $|A\rangle$ polarized incident light [Figure \ref{Figure4} (a-c)].  The simulated fields in Fig. \ref{Figure4} (d) show E-field vortices. The induced current vector moments follow the electric field around the magnetic pole [Fig. \ref{Figure4} (e)] that circulates axially from head to tail. Similarly\bluee{,} in the smaller pillar\bluee{,} the magnetic dipole circulates axially through the current (E-field) vector moments. 
This leads to the formation of a toroidal moment ($T$) perpendicular to the $z$ axis, as shown in the schematic in Fig. \ref{Figure4} (b) and (c). The near-field radiation pattern from a toroidal pole and an electric dipole is identical~\cite{Koshelev2019,Yang2019}, and the interaction between these poles form the non-radiative anapole. For the smaller elliptical pillar, we observe a dipole and a magnetic toroidal pole axially forming an anapole state. For the larger elliptical pillar, this combination of magnetic toroidal and electric multipoles forms a hybrid non-radiating anapole state [Fig. \ref{Figure4} (f)].  

%
%
Given the different eigenmodes excited for the base linear orthogonal polarizations, the modes excited for  $|L\rangle$ and  $|R\rangle$  are different, which leads to the difference in the diffraction patterns (that can be extended to any elliptical polarization state). The modal differences are highlighted further in Section S3 of the Supplemental Information.

\section{Conclusion}
In conclusion, we demonstrate a dielectric meta-polarimeter that provides a  unique diffraction pattern for any arbitrary polarization state on the Poincaré sphere. The Stokes parameters were calculated from the intensities of four spots in the diffraction pattern, yielding an experimental measurement fidelity of up to  99\%. 
The formation of the anapole state makes it possible to detect not just linear but elliptical polarization states with the same design. With this single optical component, we achieved polarimetry measurements out-performing  existing polarimetric techniques in terms of size and complexity. Although we have focused here on an experimental demonstration with pure polarization states \bluee{(by using pure input states and normalized Stokes vectors)}, 
 general polarization states--including mixed states--could be measured,  e.g., by using a maximum likelihood approach for determination of the full density matrix as done in Ref.~\citenum{Wang2018}, corresponding to the determination of the full set of Stokes parameters. This could be useful, e.g., for quantifying the degree of polarization of light or for full quantum tomography of polarization states.
\section{Acknowledgements}
The Authors would like to acknowledge the technical staff of the James Watt Nanofabrication facility for assistance in fabrication. D.F. acknowledges financial support from the Royal Academy of Engineering under the Chairs in Emerging Technologies scheme and the UK Engineering and Physical Sciences Research Council (grant EP/M009122/1). D.F. acknowledges AFOSR grant $\#$ FA9550-21-1-0312. \bluee{A.C.D. acknowledges support from the EPSRC, Impact Acceleration Account (EP/R511705/1).  M.C. acknowledges support from the UK Research and Innovation (UKRI) and the UK Engineering and Physical Sciences Research Council (EPSRC) Fellowship (“In-Tempo” EP/S001573/1). T.S.N. and D.C. acknowledge funding from the European Union’s Horizon 2020 research and innovation programme under the Marie Skłodowska-Curie grant agreement No. 765426 (TeraApps).} 



\section{Supplementary Information}
The supplementary information contains experimental results and simulations that complement the main results of the manuscript. Section S1: Lumerical FDTD simulations that show the design principle used. Section S2: The diffraction patterns obtained experimentally are compared with simulations. We also show repeatability from another fabricated device. Section S3: Contains the simulated E-H fields for $|H\rangle$, $|V\rangle$, $|A\rangle$ and $|D\rangle$ polarization states, and detailed E-H fields that show different modal resonances in $|R\rangle$ and $|L\rangle$ states.


\section{Methods}
\subsection{Fabrication}
The substrate was a borosilicate glass slide with the dimensions 20 mm × 20 mm and a thickness of 515 $\mu$m. 520$\pm$5 nm of amorphous Silicon ($\alpha$-Si) was deposited using PECVD on the borosilicate substrate that was attached to a carrier wafer (with a layer of SiO$_2$). The samples were cleaned with acetone and IPA in an ultrasonic bath. The e-beam resist (All-resist PMMA 632.06 50K) was spin coated at 4000 rpm for 60 s and baked in the oven at 180$^{\circ}$C for 30 mins. This was followed by spin coating a second layer of e-beam resist (All resist PMMA 679.02 950K) at 4000 rpm for 60 s followed by baking it in the oven at 180$^{\circ}$C for 60 mins. A 20-nm layer of Al was deposited using the e-beam evaporator Plassys II which acts as a charge conduction layer for the e-beam writing. The exposed patterns were immersed in a solution of MFCD-26 for 2 mins to remove this charge conduction layer. After thoroughly rinsing in IPA and blow dried with N2 gun, the samples were developed in 2:1 ratio of IPA:MIBK for 30 s at 23$^{\circ}$C, followed by a rinse in isopropyl alcohol (IPA) for 30 s. Plasma  ashing with oxygen in the Oxford Instruments Plasmalab 80 Plus reactive ion etcher (RIE) was done to remove any developed residue. This was followed by a metallisation step: 50 nm of NiCr deposited by the e-beam evaporator which acts as a hard mask for etching. The samples were placed in an acetone bath maintained at 50$^{\circ}$C for lift-off. The samples were etched with a C4F8/SF6 chemistry in STS ICP followed by cleaning in acetone, IPA and HMDSO (without ultrasonic). The NiCr layer was removed using a chrome etchant followed by Nitric acid (to remove the Ni layer). Finally, the samples were cleaned in acetone and IPA and imaged with the FEI Nova NanoSEM 630 (Scanning Electron Microscope). 

\subsection{Simulation Setup}
The amorphous silicon pillars defined in the unit cell were placed on a fused silica substrate. The material parameters for $\alpha$-Si were measured from the ellipsometer Bruker in the James Watt Nanofabrication Center (JWNC, University of Glasgow) clean room and the refractive index information and used in the simulations. For transmission simulation of the far-field diffracted spots, periodic boundary conditions were used around the super cell [shown in Figure \ref{Figure1} (a)]. A mesh grid with a maximum cell size of 5 nm was defined in the vicinity of the nanopillars and the interface with the substrate. Steep angle boundary conditions were used in the FDTD simulation to absorb all the light at the boundaries and prevent any spurious reflections. The metasurface was illuminated by a 808 nm plane-wave source from within the substrate, and the transmission spectra were recorded by a monitor placed on the opposite side of the nanopillars. 
A total-field scattering-field (TFSF) source was used for determination of the scattered field from the nanopillars. To obtain the $E$ and $H$ vector fields for the various polarization states a cell area was considered with perfectly matched layers (PML) boundary conditions in the $x$, $y$ and $z$ axes. The FDTD boundaries were several orders of magnitude greater than the wavelength $\lambda$. The stability factor was set to 0.7 in order to achieve simulation convergence. 

\subsection{Measurement Setup}
For the single shot imaging from the metasurface polarimetry the measurement setup is shown in Figure \ref{Figure1} (d). A Coherent Chameleon Discovery laser at 808 nm was coupled into a fiber. To set the input polarization state onto the metasurface, a polarizing beam splitter (PBS), half wave-plate (HWP) and quarter wave-plate (QWP) was used. For the collection optics of the diffraction pattern from the metasurface a combination of a condenser lens and 2” Fresnel lens was used. All images were collected with an Andor Marana sCMOS camera with an integration time of 7 ms.



\providecommand{\latin}[1]{#1}
\makeatletter
\providecommand{\doi}
  {\begingroup\let\do\@makeother\dospecials
  \catcode`\{=1 \catcode`\}=2 \doi@aux}
\providecommand{\doi@aux}[1]{\endgroup\texttt{#1}}
\makeatother
\providecommand*\mcitethebibliography{\thebibliography}
\csname @ifundefined\endcsname{endmcitethebibliography}
  {\let\endmcitethebibliography\endthebibliography}{}


%


\clearpage
\onecolumngrid
\renewcommand{\thefigure}{S\arabic{figure}}
\renewcommand{\theequation}{S\arabic{equation}}
\renewcommand{\thesection}{S\arabic{section}}
\renewcommand{\thesubsection}{s\arabic{section}}
\renewcommand{\thetable}{S\arabic{table}}
\setcounter{figure}{0}
\setcounter{equation}{0}
\setcounter{section}{0}
\setcounter{subsection}{0}
\setcounter{table}{0}

\makeatletter
{\large \bf Supplementary Material for\\``\@title "}
\maketitle

\vspace{12pt}

This supplementary information contains simulations and experimental results that complement the main manuscript text.
\section{Simulations for metasurface design} 
\begin{figure*}[!htbp]
  \centering
  \includegraphics[width=0.8\textwidth]{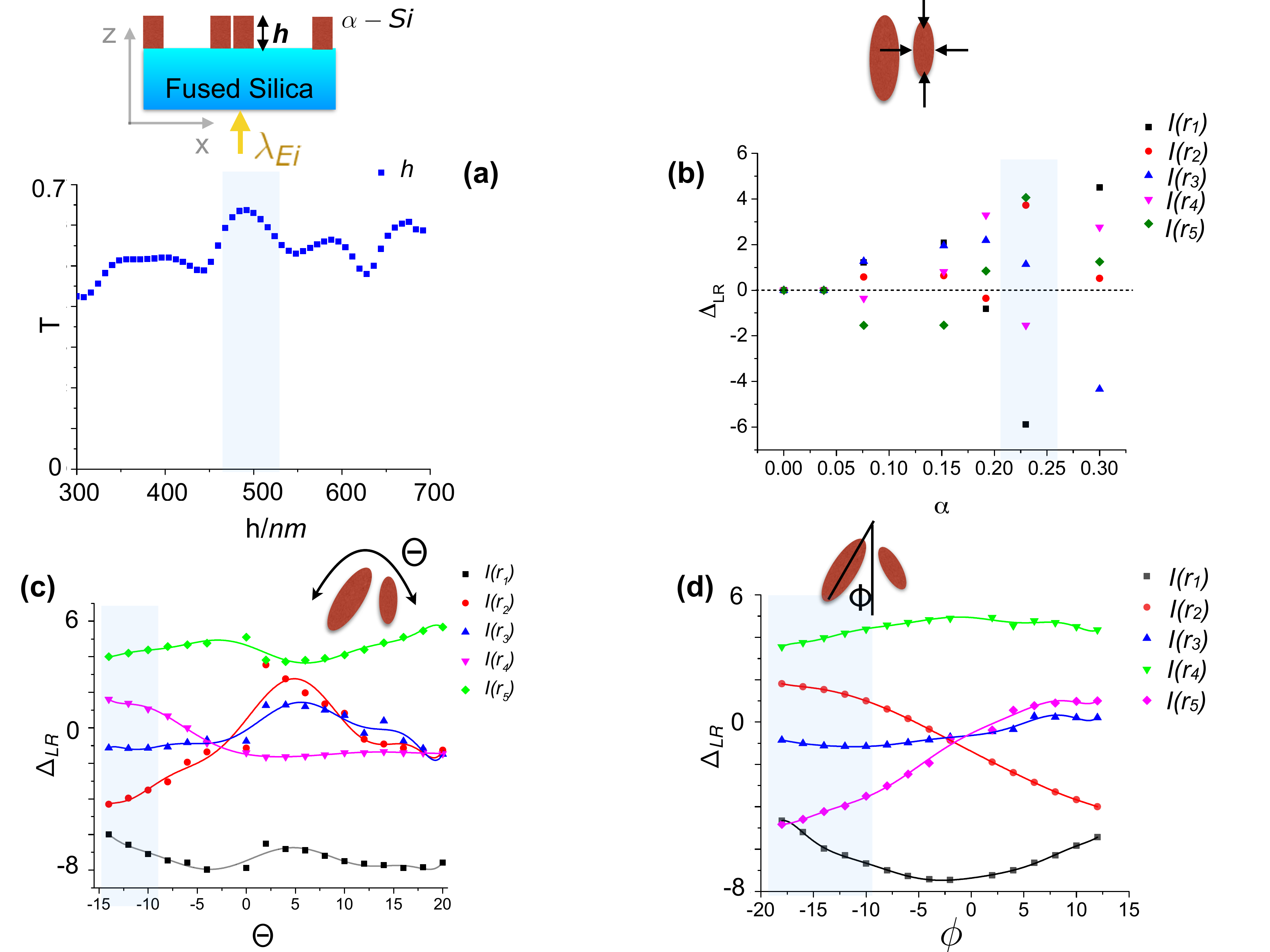}
\caption{(A) Lumerical FDTD simulations for height $h$ of the metasurface nanopillars. The height giving the highest transmission was within the range 485-525 nm. (B) Using design principles 1 and 2 for the asymmetry in geometry, the difference $\Delta_{LR}$ for the diffracted spots is shown and an ideal value of 0.25 is chosen. (C) { $\Delta_{LR}$ as a function of the orientation of the elements} (D) $\Delta_{LR}$ as a function of the orientation of the bi-atom.}
\label{FigureS1}
\end{figure*}

The geometry and orientation of the meta-atoms of the unit cell were varied until we obtained the best trade-off for the design principle: 1) highest diffraction efficiency and 2) highest difference in intensity between $|R\rangle$ and $|L\rangle$ incident polarization state (i.e., left- and right-circular polarization states, respectively). We start off with Lumerical finite-difference time-domain (FDTD) simulations on the height of the nanopillars. The shaded region in Figure \ref{FigureS1} (a) shows the ideal range in which the highest transmission efficiency ($63.7\%$) was obtained.
For the bi-meta-atom arrangement, as noted in Refs~\citenum{Leitis2019} and \citenum{Koshelev2018}, a small change in the symmetry creates leaky quasi bound-in-continuum (BIC) states. Using the dimensions and orientation of the bi-atom arrangement, we are able to tune the resonant modes through polarization of light.
In Ref.~\citenum{Liu2018}, a bi-atom arrangement that maintained symmetry along the y-axis would in principle excite the same modes for both $|L\rangle$ and $|R\rangle$ states, as there is no asymmetry in the linear polarized state to break the symmetry dependence. So $\alpha$ is the symmetry parameter in the geometry, that is how the one of the bi atoms is scaled to the other, as shown in Figure \ref{FigureS1} (b). Using the values in the window highlighted,  the angle between the meta-atoms ($\phi$) was varied and $\Delta_{LR}$ was calculated from simulation results and plotted in Figure \ref{FigureS1} (c). Finally, the orientation of the entire bi-meta-atom, $\Theta$, was varied to obtain the highest $\Delta_{LR}$.

\section{Measurements} 
\begin{figure*}[!htbp]
  \centering
  \includegraphics[width=0.8\textwidth]{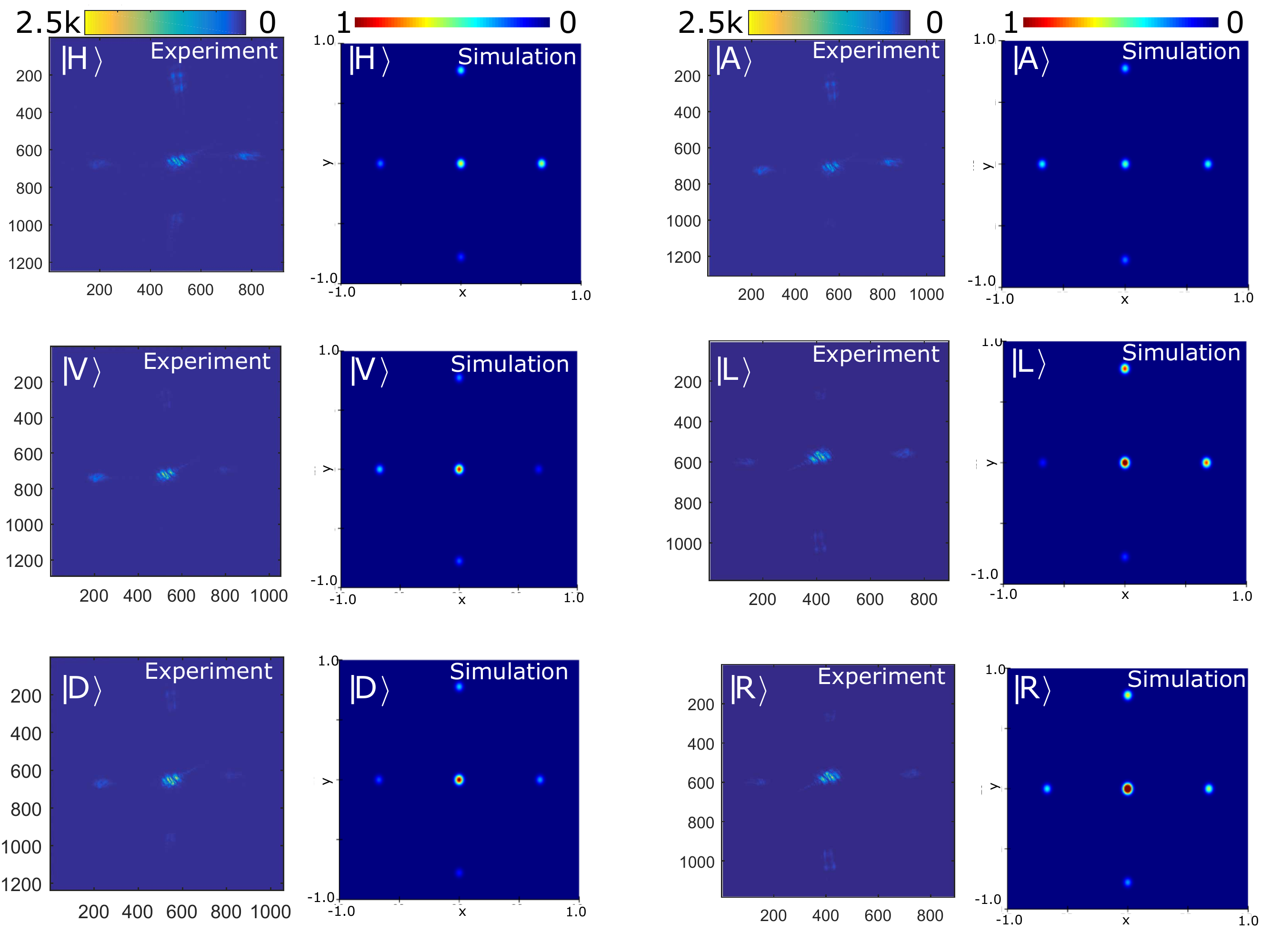}
\caption{ Diffraction pattern from a metasurface polarimeter along with the simulated results for $|H\rangle$, $|V\rangle$, $|D\rangle$, $|A\rangle$, $|L\rangle$ and $|R\rangle$ states respectively.}
\label{FigureS2}
\end{figure*}

\begin{figure*}[!htbp]
  \centering
  \includegraphics[width=0.8\textwidth]{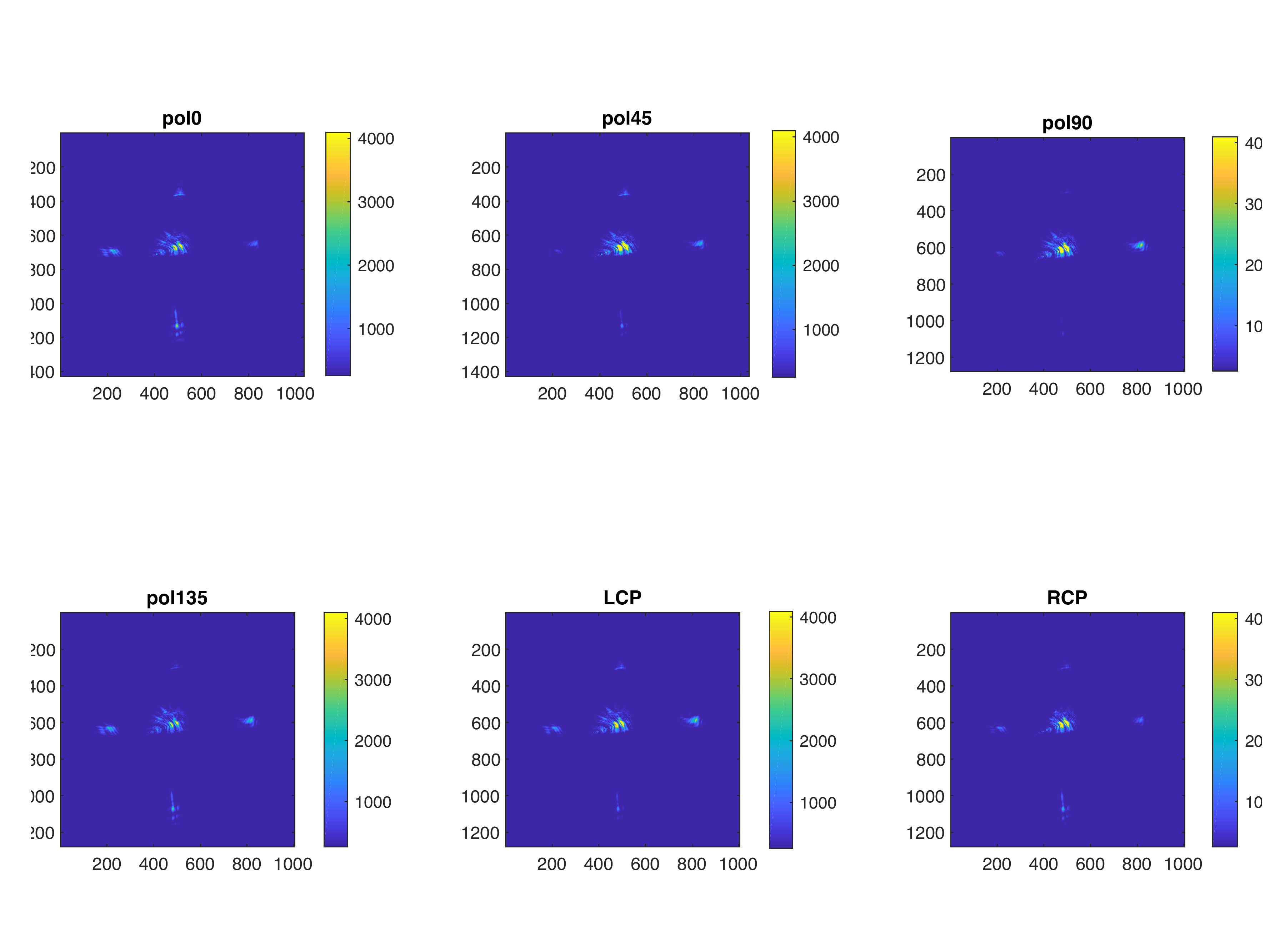}
\caption{Test of reproducibility by showing the diffraction pattern from another metasurface device.}
\label{FigureS3}
\end{figure*}

\begin{figure*}[h!]
  \centering
  \includegraphics[width=0.8\textwidth]{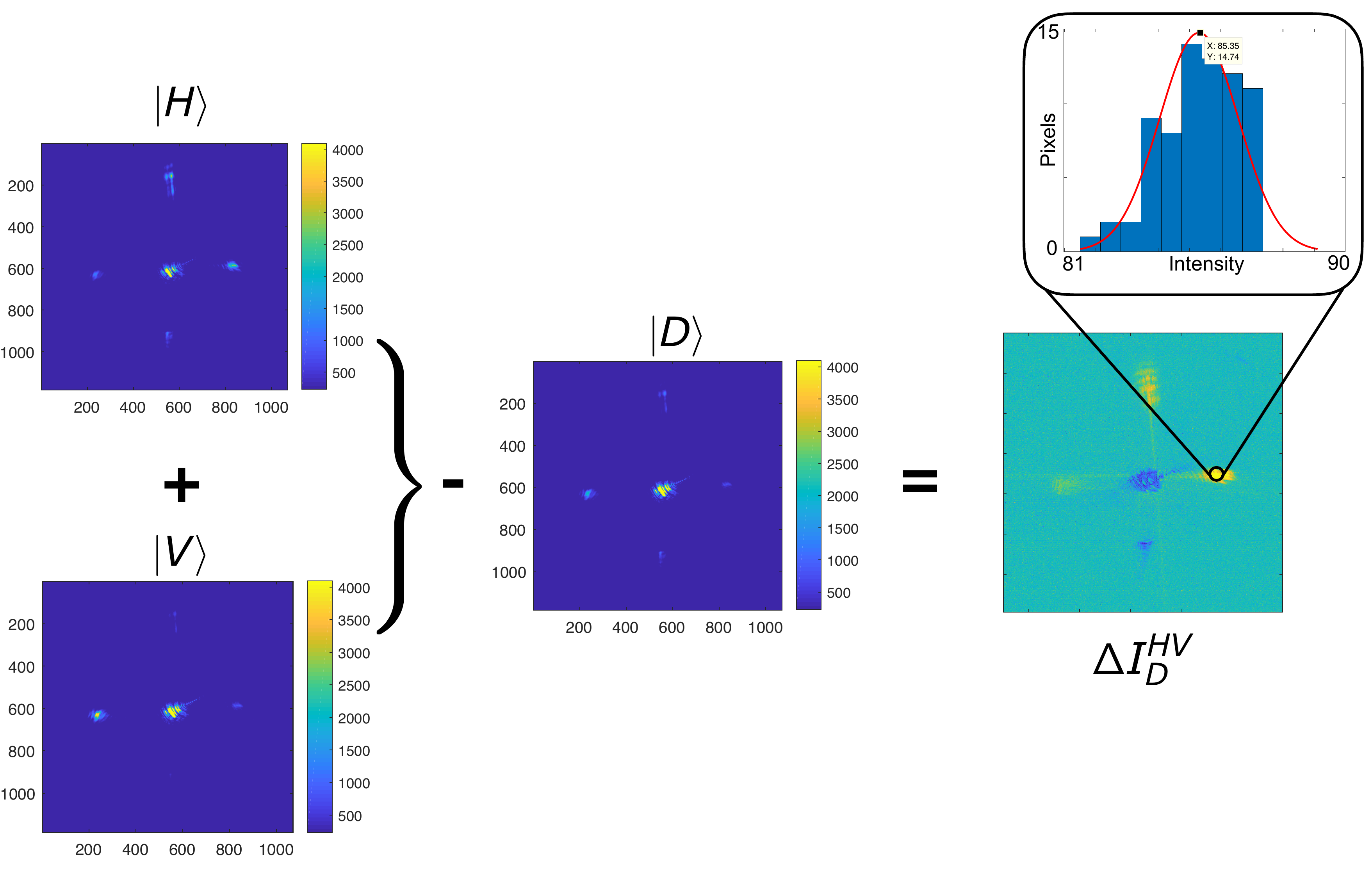}
\caption{The intensity difference patterns ($\Delta I_{D}^{HV}$)  were calculated as shown for polarization state $|D\rangle$ with respect to the combined intensities of $|H\rangle$ and $|V\rangle$. The probability distribution was obtained in 2 dimensions to get the exact value for every diffracted spot.}
\label{FigureS4}
\end{figure*}

The diffraction patterns for the orthogonal basis states and the simulated results are shown in Figure~\ref{FigureS2}. To highlight the repeatability, we show the diffraction pattern for the same polarization states from another device in Figure~\ref{FigureS3}. In the main text, we quantify the diffraction pattern by taking the difference of the diffraction pattern from an arbitrary polarization state and the combined intensities of $|H\rangle$ and $|V\rangle$ polarization states. This is shown in Figure~\ref{FigureS4} where $\Delta I_D^{HV}$ is plotted. Given the spread of the diffraction spot, we take the probability density function along the $x$ and $y$ axis. This is shown in the inset in Figure~\ref{FigureS4}.

\section{$\mathbf{E}$-$\mathbf{H}$ fields from the meta-atoms for orthogonal basis polarization states} 
\begin{figure*}[h!]
  \centering
  \includegraphics[width=0.8\textwidth]{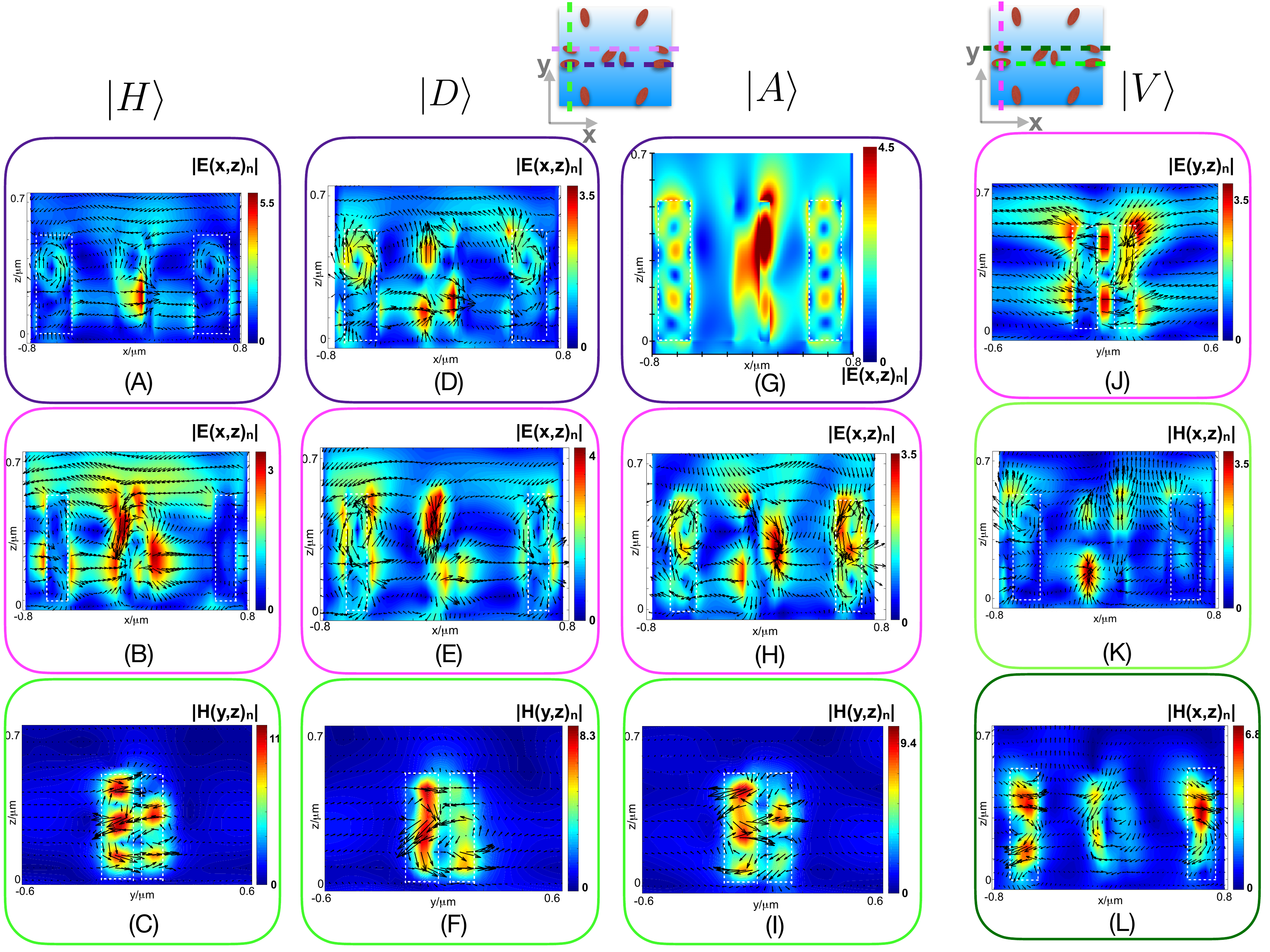}
\caption{ $|\mathbf{E}|$ and $|\mathbf{H}|$ field profiles normalized to the intensity of the incident light under $|H\rangle$, $|V\rangle$, $|A\rangle$  and $|D\rangle$  polarizations. The $|\mathbf{H}|$ fields are in green and the $|\mathbf{E}|$ fields in purple and pink. }
\label{FigureS5}
\end{figure*}
\begin{figure*}[htbp]
  \centering
  \includegraphics[width=\textwidth]{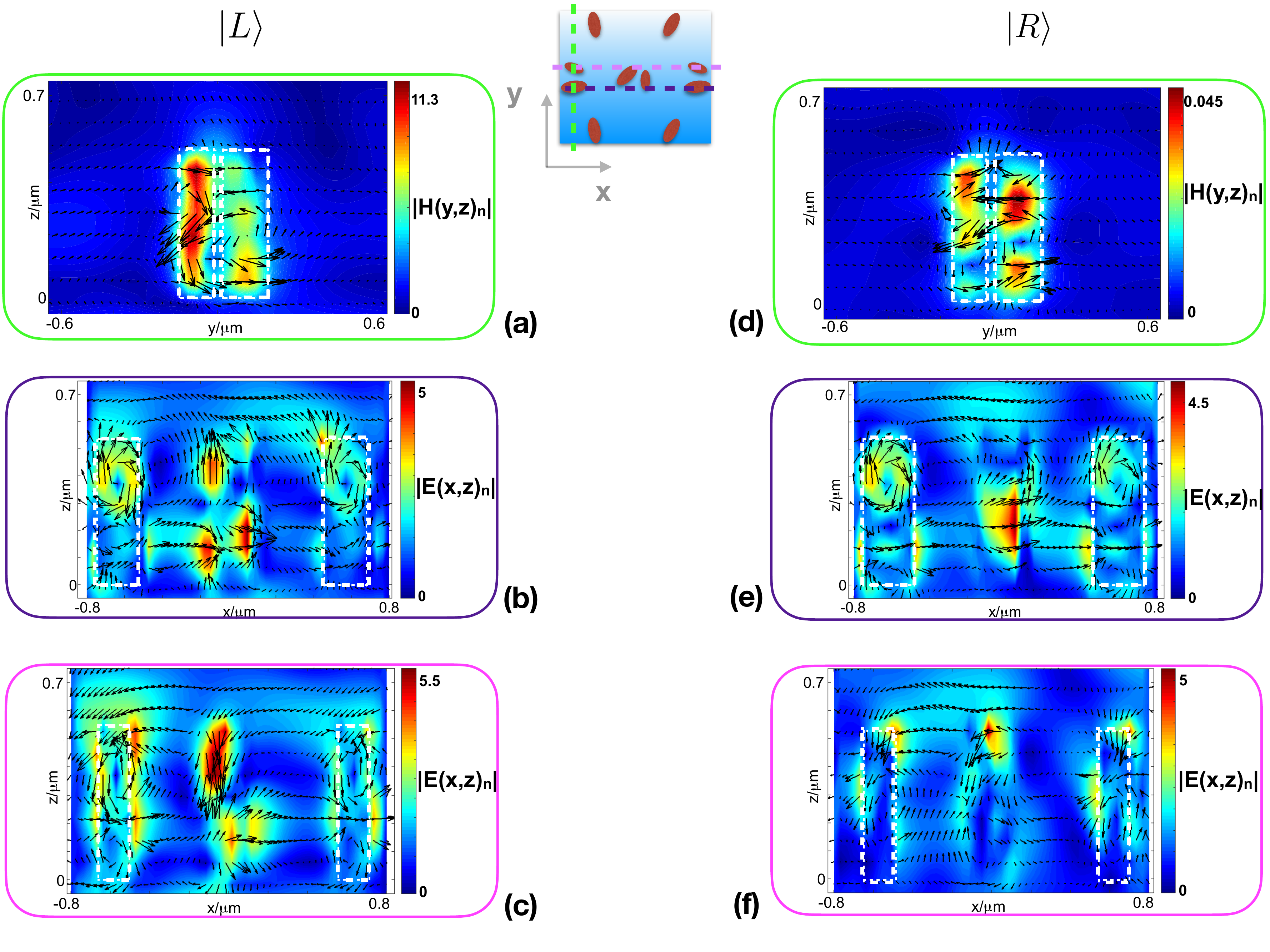}
\caption{ $|\mathbf{E}|$ and $|\mathbf{H}|$ field profiles normalized to the intensity of the incident light under $|R\rangle$  and $|L\rangle$  polarizations. The $|\mathbf{H}|$ fields are in green and the $|\mathbf{E}|$ fields in purple and pink. }
\label{FigureS6}
\end{figure*}

Figure \ref{FigureS5} shows the simulated $|E(x,z)_n|$ and $|H(y,z)_n|$ of the bi-meta-atoms arrangement as indicated in the schematic for $|H\rangle$, $|V\rangle$, $|A\rangle$  and $|D\rangle$ polarization states. For the incident $|V\rangle$ polarization state, the E-field shows an electrical resonance from the mode confined between the pillars with no excited eigenmodes within the larger pillar and a $TM_{21}$ mode oriented along the larger axis of the smaller elliptical pillar. Similarly analyzing the E-H fields for polarization $|H\rangle$ we observe a magnetic resonance in the larger elliptical pillar [$TM_{21}$ mode in Figure \ref{FigureS5} (d)] and an electrical resonance in the smaller elliptical pillar [Figure \ref{FigureS5} (e)]. 
We observe an anapole state for the $|A\rangle$ polarization state and no anapole state signature in the E-field for the $|D\rangle$ polarization state. For the $|D\rangle$ polarization state: we observe a magnetic resonance in the larger pillar [inferred from electrical vortex in Figure \ref{FigureS5} (d)] and we do not observe an electrical dipole in the smaller pillar [Figure \ref{FigureS5} (e)].
Figure~\ref{FigureS6} (a) shows the $|\mathbf{H}(y,z)n|$ field distribution for $|L\rangle$ polarized incident light wherein a more axial field is observed indicating a stronger Mie resonance than Fabry P$\acute{e}$rot mode. This is significantly different from with the $\mathbf{H}$-field for $|R\rangle$ polarized incident light which shows a stronger Fabry P$\acute{e}$rot mode, with a magnetic octupole (MO) and a magnetic dipole (MD) in the larger and smaller pillars respectively. 

\begin{figure*}[htbp]
  \centering
  \includegraphics[width=0.8\textwidth]{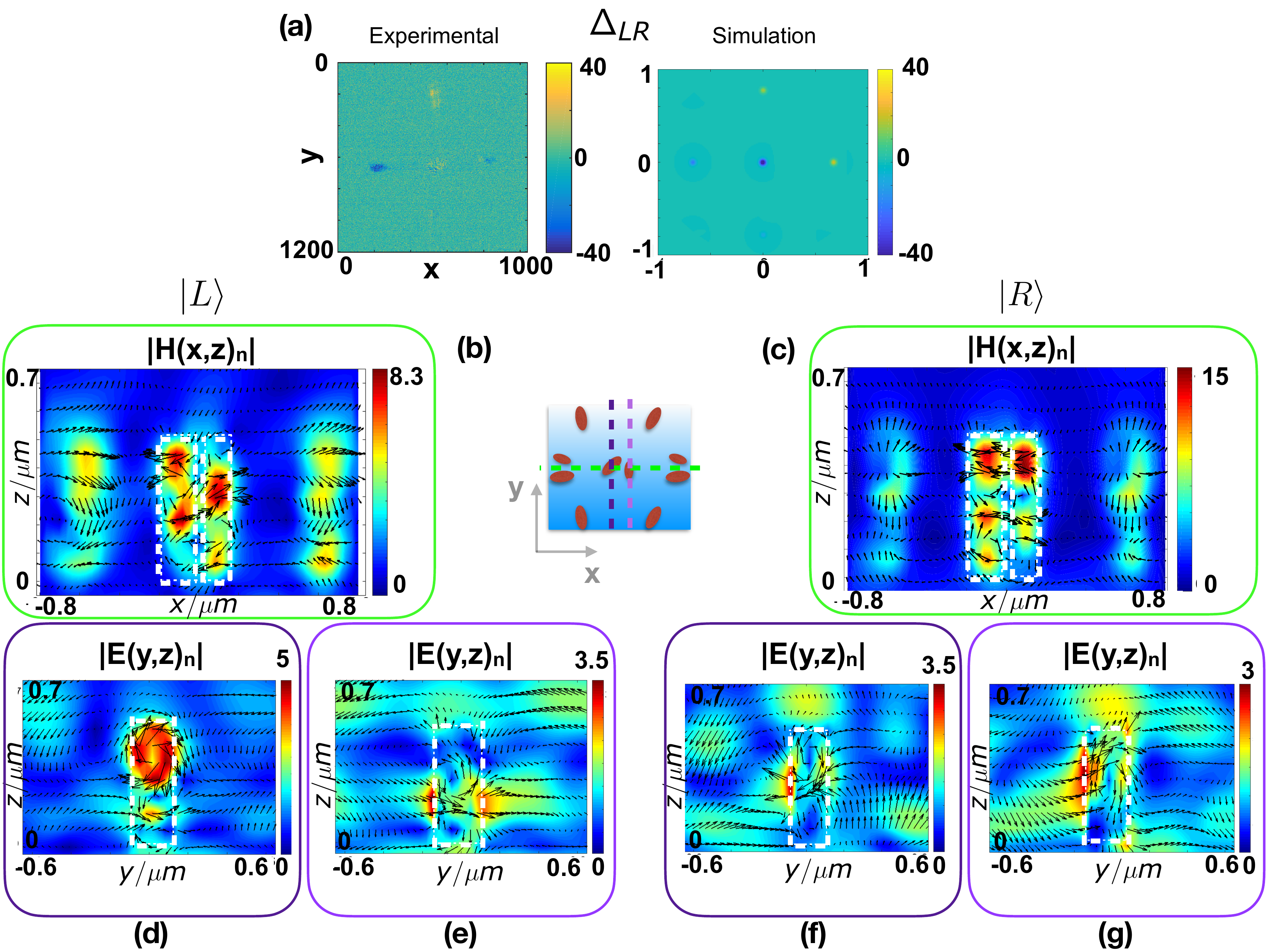}
\caption{ $|E|$ and $|H|$ field profiles normalized to the intensity of the incident light under $|R\rangle$  and $|L\rangle$  polarizations. The $|H|$ fields are in green and the $|E|$ fields in purple and pink. (A) Intensity difference, $\Delta_{LR}$, between  $|L\rangle$ and  $|R\rangle$ polarizations for each measured diffracted spot plotted against the simulated intensity difference. (B) and (C) The magnetic field $|H(x,z)_n|$ for $|L\rangle$ and $|R\rangle$ polarization states, respectively. The electrical fields are shown in (E) and (F) for $|L\rangle$ polarization and, (G) and (H) for $|R\rangle$ polarization. Due to both these radiative modes the central spot has a higher intensity for  $|R\rangle$ polarized light then  $|L\rangle$ polarized light.}
\label{FigureS7}
\end{figure*}

Figure~\ref{FigureS7} (a) (i) shows the difference in intensity of the diffraction spots between $|L\rangle$ and $|R\rangle$ polarization states, with the simulation results in (ii). Figure~\ref{FigureS7} (b) is the normalized H-field, $|H(x,z)n|$ for $|L\rangle$ polarized incident light. For this polarization of light, the simulations show a mix of a magnetic resonance and a magnetic vortex formed between the two pillars near the base. This is supported by the electrical resonance observed in Figure ~\ref{FigureS7} (f) near the base which might suggest reflection. A radiative mode is seen in with a stronger $TM_{21}$ mode mixed with a weaker $TE_{12}$ mode as shown in Figure~\ref{FigureS7} (e).  On the other hand for $|R\rangle$  polarized light from the $|\mathbf{H}(x,z)n|$ field a magnetic dipole resonance between the pillars is observed as shown in Figure~\ref{FigureS7} (c). Unlike $|L\rangle$  polarized light, an electrical resonance in the larger nano pillar [Figure~\ref{FigureS7} (g)] and a weaker E-field vortex is observed in the smaller nano pillar [Figure~\ref{FigureS7} (h)].

Although the E-field profiles in the larger pillar is similar for both polarizations as seen in Figures ~\ref{FigureS6} (b) and (c) the smaller pillar for $|L\rangle$ polarized light shows a modal interaction between a weak $TM_{21}$ and $TE_{12}$ mode whereas for $|R\rangle$ polarized light no resonant modes in the same pillar is observed. Since the $E$ and $H$ field profiles are unchanged in the extreme lattice points (single meta-atoms) for $|L\rangle$ and $|R\rangle$ polarized light, we can conclude that the difference in circular as well as elliptical polarized light are due to the different modes that couple with the incident in the bi-meta-atom arrangement.


\providecommand{\latin}[1]{#1}
\makeatletter
\providecommand{\doi}
  {\begingroup\let\do\@makeother\dospecials
  \catcode`\{=1 \catcode`\}=2 \doi@aux}
\providecommand{\doi@aux}[1]{\endgroup\texttt{#1}}
\makeatother
\providecommand*\mcitethebibliography{\thebibliography}
\csname @ifundefined\endcsname{endmcitethebibliography}
  {\let\endmcitethebibliography\endthebibliography}{}

\end{document}